# A Classical Model of Speculative Asset Price Dynamics

Sabiou M. Inoua
*Chapman University*, inoua@chapman.edu

Vernon L. Smith
*Chapman University*, vsmith@chapman.edu

# A Classical Model of Speculative Asset Price Dynamics

## Comments

ESI Working Paper 21-21

This paper later underwent peer review and was published in *Journal of Behavioral and Experimental Finance*, volume 37, in 2023.

The additional files contains the working paper version of this article. It was also previously titled "Re-tradable Assets, Speculation, and Economic Instability".

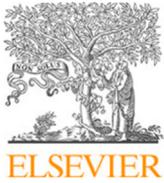
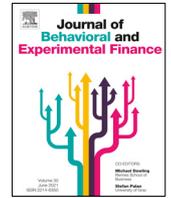

Full length article

# A classical model of speculative asset price dynamics<sup>☆</sup>

Sabiou M. Inoua [\*,1], Vernon L. Smith [1]

*Chapman University, Economic Science Institute, 1 University Dr, Orange, CA 92866, United States of America*

A B S T R A C T

In retrospect, the experimental findings on competitive market behavior called for a revival of the old, classical, view of competition as a collective higgling and bargaining process (as opposed to price-taking behaviors) founded on reservation prices (in place of the utility function). In this paper, we specialize the classical methodology to deal with speculation, an important impediment to price stability. The model involves typical features of a field or lab asset market setup and lends itself to an experimental test of its specific predictions; here we use the model to explain three general stylized facts, well established both empirically and experimentally: the excess, fat-tailed, and clustered volatility of speculative asset prices. The fat tails emerge in the model from the amplifying nature of speculation, leading to a random-coefficient autoregressive return process (and power-law tails); the volatility clustering is due to the traders' long memory of news; bubbles are a persistent phenomenon in the model, and, assuming the standard lab present value pattern, the bubble size increases with the proportion of speculators and decreases with the trading horizon.

© 2022 The Author(s). Published by Elsevier B.V. This is an open access article under the CC BY-NC-ND license (http://creativecommons.org/licenses/by-nc-nd/4.0/).

## 1. Introduction

Laboratory market experiments have established the stability and efficiency of competitive markets organized notably under the double-auction trading institution (Smith, 1962). The experimental findings (reviewed, e.g., in Plott, 1982; Smith, 1982; Smith and Williams, 1990; Davis and Holt, 1993; Holt, 1995, 2019) challenge core tenets of neoclassical value theory (the requirement for large number of traders, market clearance, complete information of supply and demand, and most notably passive price-taking behaviors, which bypass the central problem of price discovery) and, in retrospect, they call for a theory of competitive markets rooted in the old, classical, view of competition as a collective higgling and bargaining process, founded on reservation prices, as the authors' recent reappraisal of the experiments and the classics of value theory suggests (Inoua and Smith, 2020b,c,a, 2021, 2022a,b,c). The stability and efficiency of lab markets, which are robust to various supply and demand conditions, do not hold, however, for a good that is retradable for capital gains (Dickhaut et al., 2012; Inoua and Smith, 2022c), for then the stabilizing virtue of competition is counteracted by speculation.

In this paper, we specialize the classical methodology to deal with speculation, thus complementing the theory of competitive markets for non-retradable goods (Inoua and Smith, 2021) with a model of a speculative asset market that assumes typical features of a field or lab market, classical micro-foundations (with the state of risk aversion in the market completely specified by a distribution of minimum acceptable rate of return), and adaptive expectations.[2] We model traders, endowed with initial holdings of cash and asset units, who compete to trade asset units

---

☆ We thank J. Huber for initiating the experimental project, "Nobel and Novice", for which an earlier draft of this paper, then titled "Re-tradable Assets, Speculation, and Economic Instability", was submitted to the appreciation of many referees. We benefited from the reviews of 500+ referees, whom we thank. This revised version is centered on the model of speculative price dynamics, emphasizing the classical micro-foundations of the model. Other topics discussed in the previous draft (such as asset retradability) are better discussed in specialized works; see Inoua and Smith (2022c). We thank J. Huber, R. Kerschbamer, C. König, for stimulating comments and valuable suggestions, and S. Palan (Editor) for the awesome task of summarizing 500+ referee reports and providing very valuable suggestions. To all these acknowledgments, the usual disclaimer applies. We thank the Charles Koch foundation for financial support. Finally, we dedicate this paper to the memory of Clas Wihlborg.

\* Corresponding author.

*E-mail addresses:* inoua@chapman.edu (S.M. Inoua), vsmith@chapman.edu (V.L. Smith).

[1] The authors contributed equally to the article.

[2] Briefly, by the classical approach we mean adopting a realistic approach to individual behaviors and interactions and deriving economic regularities as collective patterns emerging from these behaviors and interactions. For a detailed discussion of the classical methodology, as we reappraise it, see, again, Inoua and Smith (2020b,c, 2022b).

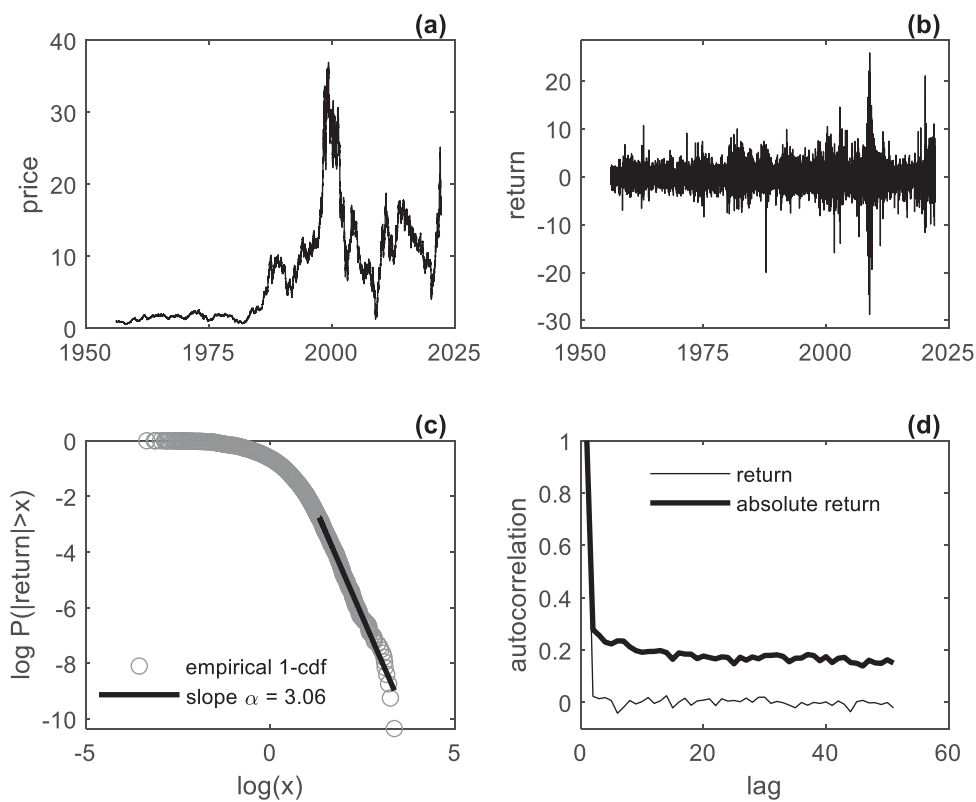

**Fig. 1.** Ford Motor Company stock: (a) price; (b) return (in percent); (c) cumulative distribution of volatility in log–log scale, and a linear fit of the tail, with a slope close to 3; (d) autocorrelation function of return, which is almost zero at all lags, while that of volatility is nonzero over a long range of lags.[5]

based on their news-corrected adaptive expectations of future price change or asset mispricing, and are willing to trade for at least a minimum acceptable return. Specific predictions of the model can be derived that can be tested experimentally. At this point, however, our goal is primarily theoretical: we develop the theoretical foundations of the model and use its linearized version to explain three general stylized facts of speculative asset prices, well-established both empirically and experimentally: their excess, fat-tailed, and clustered volatility.[3]

The excess volatility puzzle (Shiller, 1981; LeRoy and Porter, 1981), a challenge to the efficient-market hypothesis (Fama, 1970), is now an equally well-received hypothesis (as attests the Nobel prize of 2013); moreover, asset experiments (Smith et al., 1988) provide unambiguous evidence to this hypothesis, by allowing complete control over the fundamental value (for reviews, see Porter and Smith, 2003; Palan, 2013). The fat tails of asset returns imply that financial volatility is more extreme than the Gaussian distribution commonly assumed: more precisely, the empirical distribution of returns is a power law with an exponent often close to 3 (Guillaume et al., 1997; Gopikrishnan et al., 1999; Plerou et al., 1999; Gabaix et al., 2006).[4]

Volatility clustering means that high-amplitude price changes tend to be followed by high-amplitude price changes, and low-amplitude price changes, by low-amplitude price changes, implying a long-memory volatility process and a nontrivial predictability in price changes, whose sign is serially uncorrelated but whose amplitude (absolute value) is long-range correlated (Bollerslev et al., 1992; Ding et al., 1993; Granger and Ding, 1994; Comte and Renault, 1996). Both the fat tails and the clusters of volatility (illustrated in Fig. 1) are also general properties of speculative prices, applying to various assets (commodities, stocks, exchange rates, options, indices), across various time scales (from a few minutes to a few weeks), on different market places, and are also confirmed in asset experiments (Kirchler and Huber, 2007, 2009). By their generality, these stylized facts should constrain any realistic theory of speculative markets.

The neoclassical approach to value theory faces foundational difficulties, the most important of which being the above-mentioned problem of price formation.[6] In finance, more specifically, these problems are compounded by the no-trade, or more precisely no-speculation, theorems, which uncover an inherent difficulty of modeling speculative trade itself in terms of expected-utility maximization and rational expectations (Rubinstein, 1975; Milgrom and Stokey, 1982; Tirole, 1982; Gizatulina and Hellman, 2019).[7]

---

[3] Throughout this paper, "volatility" refers to the magnitude (absolute value) of asset returns (percent price change), although the word is used more commonly when this magnitude is averaged across time intervals.

[4] Regarding the fat tails, Mandelbrot more precisely found a power-law exponent $\alpha < 2$ and conjectured that log-price changes follow a stable distribution, a hypothesis confirmed early by Fama (1963). But subsequent works (cited in the text) based on more extensive data found instead a power law with $\alpha \approx 3$. For a review of the financial stylized facts, see, e.g., Cont (2001, 2007), Chakraborti et al. (2011a), and Lux and Alfarano (2016).

[5] Data source: Center for Research in Security Prices (CRSP); data accessed through Wharton Research Data Services (WRDS).

[6] Despite a few able attempts at articulating more realistic neoclassical price and trade formation processes than the tatonnement story, the core problems remain unsolved. As states a review of these models: "we shall have to conclude that we still lack a satisfactory descriptive theory of the invisible hand". Hahn (1982, p. 746) More recently: "we do not have an adequate theory of value, and there is an important lacuna in the center of microeconomic theory. Yet economists generally behave as though this problem did not exist". Fisher (2013, p. 35).

Ingredients for an alternative to the elegant neoclassical approach to finance are scattered in various studies: for example, behavioral finance emphasizes various cognitive biases and other "anomalies" of the neoclassical model of rational behavior (Barberis and Thaler, 2003); and various field and lab data suggest that traders in practice follow simple adaptive expectation heuristics (Smith et al., 1988; Haruvy et al., 2007; Chow, 2011; Lahav, 2011; Anufriev and Hommes, 2012; Greenwood and Shleifer, 2014; Colasante et al., 2017; Hommes, 2021), to which come down trend-following trading strategies based on moving averages of past prices or returns (Baltas and Kosowski, 2013; Lempérière et al., 2014; Zakamulin, 2014; Levine and Pedersen, 2016; Beekhuizen and Hallerbach, 2017).[8] Moreover, theoretical and statistical models exist that combine these ingredients, notably agent-based financial models (reviewed, e.g., by Samanidou et al., 2007; Chakraborti et al., 2011b; Lux and Alfarano, 2016), which mimic the stylized facts through a mix of nonlinear mechanisms such as traders switching between trading strategies (Lux and Marchesi, 1999, 2000), which, however, are not needed for the emergence of the stylized facts in the lab (Kirchler and Huber, 2007, 2009).[9] But although models abound in this field, there is yet to emerge a general, unifying, relatively simple, micro-founded, theoretical framework that would qualify as a standard model of the stylized facts.

We propose to this end the classical model above-described, pinning down the stylized facts to their simplest causes through linear mechanisms (assuming both speculators and value-investors similarly to the agent-based literature, but not the switching between the trading strategies).[10] The fat tails emerge in our model from the intrinsic self-reinforcing nature of speculative trading: a speculative asset return forms a random-coefficient autoregressive process from which emerge power-law tails by an important theorem (Kesten, 1973).[11] The destabilizing role of speculation is intuitive and familiar, and invoked in various forms and under different names in many models (e.g., Cutler et al., 1990; De Long et al., 1990); we show here that speculation and adaptive expectations lead intrinsically to the power law of returns in a competitive market. Volatility clustering is due to the long memory that traders have of exogenous news as reflected in their expectations, if modeled as forming a nearly integrated news-corrected adaptive expectation process (to allow for long memory); an indication of this explanation is the timing of big volatility clusters, concomitant with major economic events, such as uncertainty ensuing a crisis (Fig. 2), an intuition that accords well with earlier hypotheses, theoretical (Andersen and Bollerslev, 1997) and experimental (Kirchler and Huber, 2007, 2009).[12] In fact, underlying most models of volatility clustering we analyzed, is a simple, integrated, long-memory, process usually at work amid a complex mix of mechanisms, and which, upon scrutiny, might be the ultimate cause of the phenomenon in these models: thus, the random walk of fundamental value assumed in agent-based models ([from seminal work by Lux and Marchesi, 1999, onward); the integrated GARCH model (Engle and Bollerslev, 1986), which, among the GARCH family (Engle, 1982; Bollerslev, 1986; Bollerslev et al., 1992), fits best the empirical data [but has a power law tail exponent of 2 (Mikosch and Starica, 2000, 2003)]; and similar models (e.g., LeBaron, 2001).

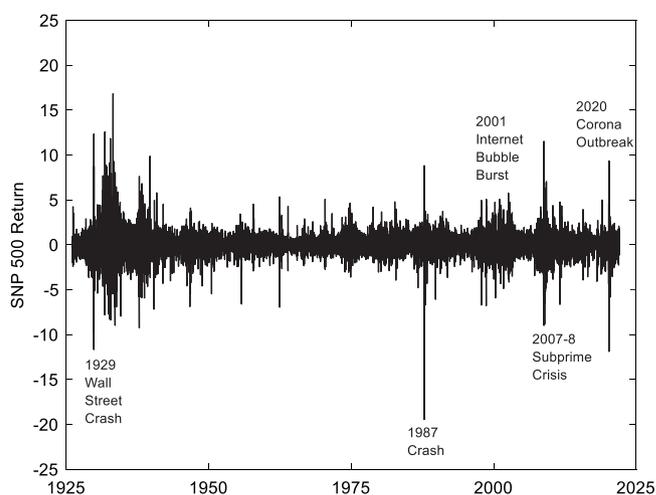

**Fig. 2.** Big volatility clusters triggered by major events (crises).[13]

The rest of the paper is organized as follows. Section 2, which starts with a few basic definitions, restates more formally the three stylized facts of asset returns. Section 3 presents the model's general setup and assumptions; Section 4 specializes the model to a return process driven by speculators only and thus explains the power law of return; and Section 5 adopts a more general specification of the model to explain the other stylized facts as well. Directions for future research are indicated in the summary and concluding remarks in Section 6.

---

[7] The stylized facts apply to speculative prices. A laboratory asset experiment uncovers nonspeculative bubbles (Lei et al., 2001), namely bubbles occurring absent asset re-trading (a necessary condition for speculation), and due, at least partly, to decision error or confusion, or the need for subject-traders to trade, the only game available. For a recent reappraisal of this "counterexample" to the speculative bubble hypothesis, reaffirming the centrality of speculation in the seminal asset market design (Smith et al., 1988), see the working paper by Tucker and Xu (2020).

[8] A moving average of a variable and an adaptive expectation of that variable are essentially the same concept mathematically. Thus, modeling typical trend-following trading strategies amounts to assuming adaptive expectations of prices, or, better, returns, as the trading strategies seem to be better formulated in terms of moving averages of past returns than moving averages of past prices (Beekhuizen and Hallerbach, 2017).

[9] The classical method, as above-defined (Footnote 2), has a few themes (realism of assumption, simple individual behaviors, emergence of complex aggregate patterns) that are echoed in the contemporary "complex systems" approach to the economy more generally, to which physicists have greatly contributed, usually applying models and techniques from statistical physics to unravel the subtle statistical microstructure of order-book-driven financial markets and the financial stylized facts in particular, a trend often known as "econophysics" (overviewed, e.g., in Mantegna and Stanley, 1999; Voit, 2003; Samanidou et al., 2007; Bouchaud, 2011; Chakraborti et al., 2011b,a; Lux and Alfarano, 2016; Bouchaud et al., 2018).

[10] The linearized version of the model invoked in this paper to explain the stylized facts appears in a previous work (Inoua, 2020), without the classical micro-foundations and the mathematical justifications that we provide here.

[11] For a complete overview of the theory of random-coefficient autoregressive (RCAR) or Kesten processes, see the monograph by Buraczewski et al. (2016), to which we often refer the reader in this paper. RCAR processes are more general and more realistic versions of the more usual ARMA processes, having, moreover, the fascinating property of allowing the emergence of a fat-tailed output from light-tailed inputs. RCAR processes already play a central, if perhaps

unfamiliar, role in finance. Thus, GARCH processes belong to this class of processes; a few theoretical models also invoke a first-order RCAR process, e.g.: the linear approximation of some agent-based models (Sato and Takayasu, 1998; Aoki, 2002; Carvalho, 2004), or a 'rational bubble' model assuming a random discount factor (Lux and Sornette, 2002), which generates a tail exponent smaller than 1. The mathematics of RCAR process is rather involved and requires advanced techniques of probability theory; in the TeXFolio:appA Appendix we offer a simple and intuitive derivation of their power-law tail behavior, in the one-dimensional case relevant for our purpose in this paper.

[12] The arrival of news was suggested as the most important cause of the fat tail and the clustering of volatility in the experiments (Kirchler and Huber, 2007, 2009).

[13] Data source: CRSP.

## 2. The three stylized facts of speculative prices

Consider (discrete) trading periods (say days) $t = 1, \ldots, T$, and include $t = 0$ merely to date the initial positions of the variables involved (rather than as a trading time). Let $\{p_t\}$ be the (adjusted) closing prices of a speculative asset XYZ at the end of each period $t = 0, 1 \ldots, T$ and let $\{d_t\}$ be the dividends paid, say, at the closing of each period $t$.

Excess volatility simply means that $p_t - v_t^e$ is persistently nonzero, where $v_t^e$ is the asset's present fundamental value, as anticipated by the market. Let $v_t$ be the asset's unknown present value, the discounted sum of (unknown) future dividends $\{d_{t+k} : k \geq 1\}$. Assuming, just for notational simplicity, a constant discount factor $(1 + \rho)^{-1}$, the asset unknown present value satisfies the stochastic recurrence equation:[14]

$$v_t = (1 + \rho)^{-1}(v_{t+1} + d_{t+1}), \tag{1}$$

whose usually considered forward-looking solution is

$$v_t = \sum_{k=1}^{\infty} \frac{d_{t+k}}{(1+\rho)^k}. \tag{2}$$

The (expected) fundamental present value is commonly defined as $v_t^e = \mathbf{E}_t(v_t)$, an optimal forecast of (2) given available information at time $t$, which in a strong version of rational expectations includes the dividend generating process itself, which we will denote generically as $G$. In the seminal asset experiments (Smith et al., 1988), intended to approximate the latter version of rational expectations (actually a stronger version: common knowledge of rational expectations), the experimenter announces publicly the dividend generating process, leading, in the standard case (Smith et al., 1988) of dividends $\{d_t\}$ randomly drawn from a known (usually a discrete uniform) distribution, to a fundamental value process declining to zero, a step function whose linear approximation is

$$v_t^e = (T - t + 1)\mathbf{E}(d_t), \quad t = 1, \ldots, T, \tag{3}$$

where here and throughout $\mathbf{E}$ is an (unconditional) expectation operator.

In this paper, we will adopt a different definition of an asset present value than the neoclassical one, $v_t^e = \mathbf{E}_t(v_t)$, based on rational expectations of the forward-looking solution of the stochastic recurrence equation (1).[15] We will assume an adaptively formed expected present value estimated from the dividend history [Eq. (14) below].

Define the asset return (or relative price change) during each period as

$$r_t = \frac{p_t - p_{t-1}}{p_{t-1}}, \quad t = 1, \ldots, T.$$

The empirical power law of speculative returns reads formally:

$$\text{prob}\{|r| > x\} \sim cx^{-\alpha}, \quad x \to \infty \ (\alpha, c > 0), \tag{4}$$

where $\alpha$, the key parameter, is called the Pareto index or tail exponent (typically close to 3) and $c$ is merely a norming constant.[16]

Throughout this paper, we estimate the power-law exponent $\alpha$ by the following least-square estimator:

$$\hat{\alpha} = \frac{\text{cov}(\log x, \log H(x) | x \geq x_{\min})}{\text{var}(\log x | x \geq x_{\min})}, \text{ where } H(x) = \text{prob}\{|r| > x\}, \tag{5}$$

and the cutoff $x_{\min}$ is optimally chosen by an algorithm by Clauset et al. (2009).[17] Graphically, the estimator (5) is simply the slope of the linear least-square fit of the (complementary) cumulative distribution of returns beyond the cutoff $x_{\min}$.

We mention in passing that (4) is more precisely the tail probability of return conditional on positive prices: by definition of the return as a ratio involving the inverse price, we should consider time periods involving nearly complete price crashes, formally the density function of price, say $f$, is positive at zero, in which case we have[18]

$$\text{prob}\{|1/p| > x\} = \int_{-1/x}^{1/x} f(z)dz \sim \frac{2f(0)}{x}, \quad x \to \infty, \tag{6}$$

a power-law tail with exponent $\alpha = 1$, which the return inherits as a product involving this power law.[19] Thus one would expect a tail exponent $\alpha \approx 1$ in asset experiments involving the declining present value (3) and exhibiting bubble-and-crash phenomenon, since $v_t^e \to 0$ implies $p_t \to 0$ in a market involving a sufficiently large number of fundamental-value investors, who tend to bring the price near its fundamental value by definition (by arbitraging away asset mispricing): this basic theoretical prediction seems to be indeed the case in experimental data, as Fig. 4 suggests, where, for statistical significance, we pooled data across experimental sessions. (See Fig. 3.)

Thus, to achieve more usual power law of returns in the lab requires implementing a more realistic pattern of fundamental value process, as did Kirchler and Huber (2007, 2009), using a random walk of dividends: see Fig. 4. Ideally one would also wish to have sufficiently long enough trading periods under homogeneous treatments (to ensure an invariant return-generating process) for statistical significance (the estimation of fat tails requires more data than that of light tails).[20]

Volatility clustering means that the magnitude (usually absolute return) of speculative returns has a long memory, in the sense of having a slowly decaying autocorrelation function:

$$\text{cor}(|r_t|, |r_{t+h}|) > 0 \text{ across many lags } h. \tag{7}$$

Formally long memory requires the ACF to decay so slowly as to be nonintegrable, meaning

$$\sum_{h=0}^{\infty} |\text{cor}(|r_t|, |r_{t+h}|)| = \infty. \tag{8}$$

Autoregressive processes (RCAR) typically exhibits exponentially decaying memory of the past: that is, under technical but general

---

[14] Here for conceptual clarity, we distinguish the asset's fundamental value, the discounted sum of future dividends $v$, an unknown (random) variable, and the asset's fundamental value $v^e$, as forecasted by the market participants. It is more common to call fundamental value the latter concept, however, namely the rational valuation (expectation) of the former variable by a representative trader.

[15] On the distinction between the two solution concepts for a stochastic recurrence equation, forward-looking (noncausal) versus backward-looking (causal), perhaps first emphasized in Vervaat (1979, Theorem 2.1), see Buraczewski et al. (2016, p. 16).

[16] The notation $f \sim g$ means that $f(x)/g(x) \to 1$ as $x \to \infty$.

[17] For the code generating the cutoff (and an alternative estimator described next), see https://aaronclauset.github.io/powerlaws/. We prefer to use the estimator (5) to the more common maximum likelihood estimator $\hat{\alpha} = n/\sum_{i=1}^{n} \log(x_i/x_{\min})$, which yields similar but slightly less accurate estimates (in terms of downward bias), as we notice based on simulations.

[18] For a sufficiently long-time horizon, a price crash is an almost sure phenomenon whenever the average return is small enough.

[19] Formally, assume $\Delta p$ is such that $0 < \mathbf{E}|\Delta p| < \infty$ and is independent from $p$, whose density function is $f$. Then (6) and Breiman's lemma (Breiman, 1965, Proposition 3; see Buraczewski et al., 2016, p. 275) implies that $\text{prob}\{|\Delta p/p| > x\} \sim 2f(0)\mathbf{E}|\Delta p|/x, x \to \infty$.

[20] One might also consider running lab experiments in which trade is uninterrupted, if cross-period leaning of subjects is not of special interest. We thank D. P. Porter for this observation.

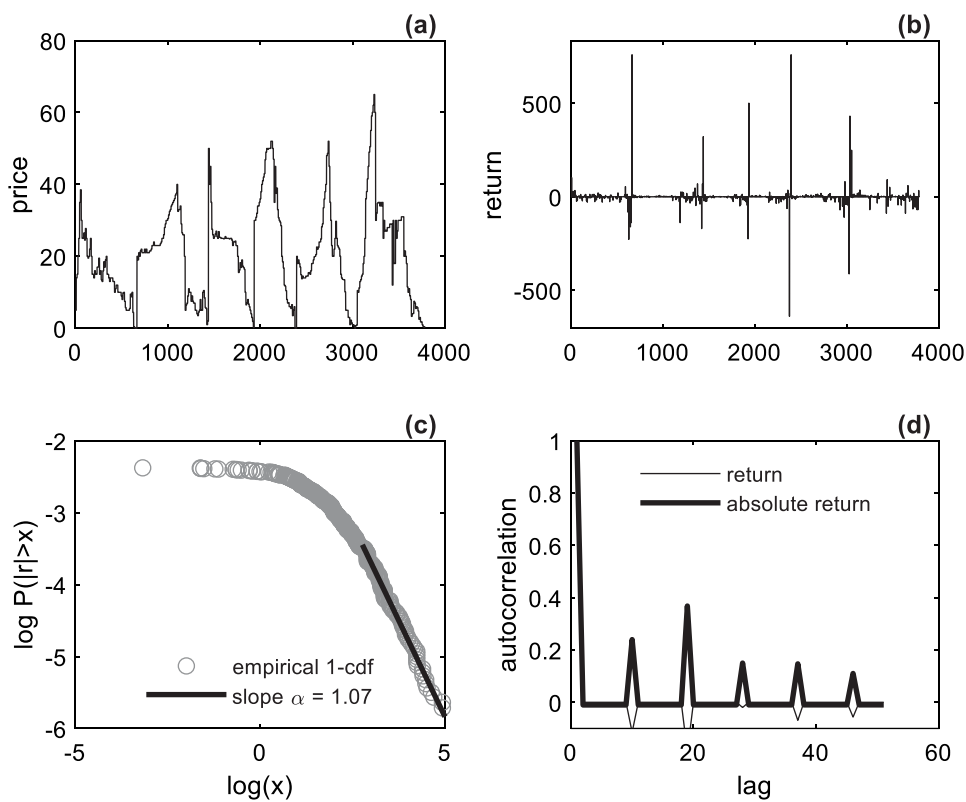

**Fig. 3.** Power law behavior in an experimental bubble-and-crash experimental data: 6 market-session data pooled. Data source: Lahav (2011).

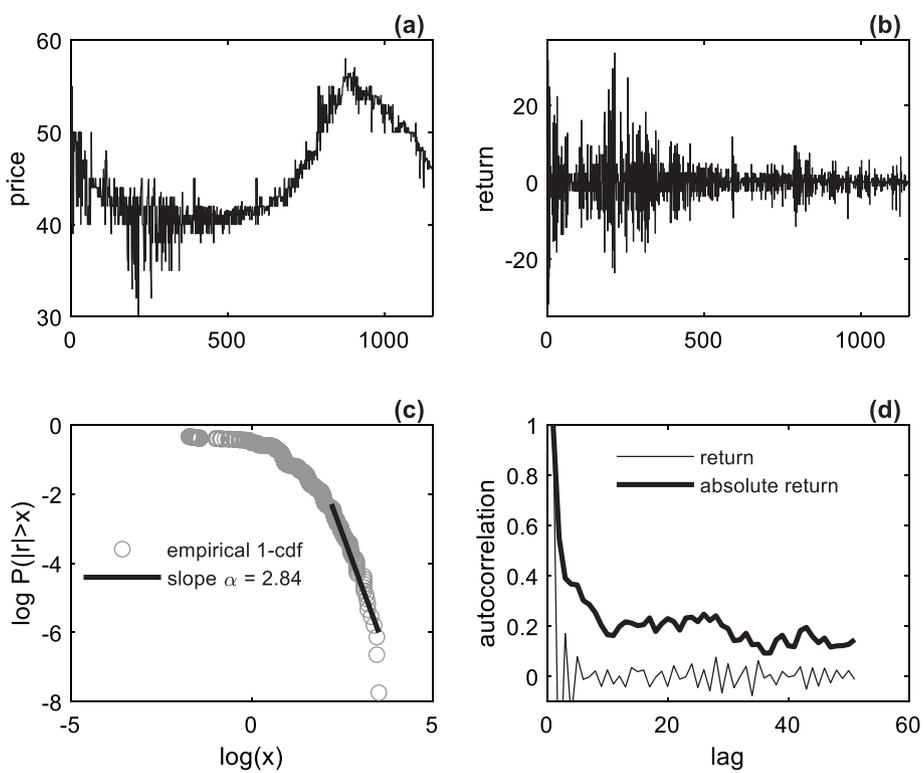

**Fig. 4.** An experimental asset market price data showing the fat tails and clusters of volatility. Data source: Kirchler and Huber (2009, Market 2, Treatment 1).

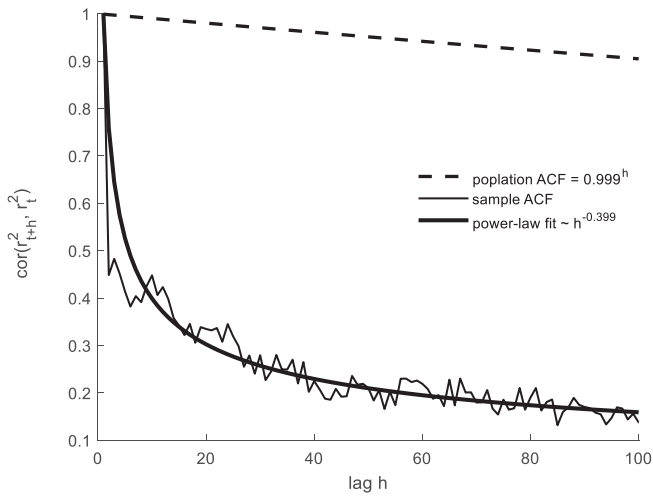

**Fig. 5.** Population versus sample ACFs for a nearly integrated GARCH (1, 1) process with $c = 0.1$, $a = 0.1$, $b = 0.899$, $a + b = 0.999$.

conditions (Buraczewski et al., 2016, p. 23) one can show that for a measurable function $f$,

$$\text{cov}(f(r_t), f(r_{t+h})) \leq c_0 \mu^h, \quad 0 < \mu < 1, \quad h \to \infty, \quad (9)$$

which is a short-memory process, except in the critical (limiting) case $\mu \to 1$. The critical case $\mu \to 1$ is, we believe, the most natural and plausible explanation for volatility clustering in terms of long memory of news (as our model suggests), despite dominant belief in a power-law model for the long memory for volatility:

$$\text{cor}(|r_t|, |r_{t+h}|) \sim h^{-\beta}, \quad 0 < \beta < 0.5, \quad (10)$$

which seems to better model the empirical ACF of volatility (for a brief review, see, e.g., Lux and Alfarano, 2016, p. 5), but which does not necessarily imply that the true (population) ACF itself follows a power law, as the following simulation based on an almost integrated GARCH (1, 1) suggests (our model exhibits the same pattern): $r_t = \sigma_t \varepsilon_t$, where $\sigma_t = a + b\sigma_{t-1} + c\sigma_{t-1}^2$, where $\{\varepsilon_t\}$ are i.i.d. (independent and identically distributed) draws from a normal distribution, $c > 0$, $a, b \geq 0$; the GARCH(1, 1) is known (Bollerslev, 1986, Equation 14) to have $\text{cor}(|r_t|^2, |r_{t+h}|^2) = (a+b)^h$, in our example $(0.999)^h$. Yet the empirical ACF seems to be better fit by a power law, as the simulation in Fig. 5 shows:

## 3. The model: General setup and assumptions

Assume:

1. (Assets) A market involving a risky asset (say a stock) that pays exogenously generated dividends $\{d_t\}$, and cash money (the means of payment) that pays $r_f$ percent interest per period. Each trader starts with initial cash and asset holdings, in total $C_0$ and $S_0$, all traders included.
2. (Traders) The market is populated by two types of traders: Type I traders, who trade based on capital gain forecasts, and Type II traders, who trade based on mispricing forecasts. Each trader $i \in I \cup II$ is willing to trade for at least a minimum acceptable return $\rho^i \geq 0$, distributed among each trader Type according to $F^J(x) = \text{prob}\{\rho^J \leq x\}$, $J = I, II$.
3. (Expectations) The traders' expectations are adaptive and news-correcting.
4. (Competition) The asset prices $\{p_t\}$ emerge competitively, by the law of supply and demand, simplified into a linear response of the asset return $r_t$ to excess demand $Z_t$.
5. (Distributions) Each unknown distribution of interest (see below) is modeled by a maximum entropy one (that is, by an uninformative prior: uniform, exponential, or normal).[21]

Let the dividend-generating process be generically $d_t = G(d_{t-1})$, $t = 1, \ldots, T$, $d_0 = 0$, where $G$ is a stochastic function that we will specialize in accordance with lab implementations: random draws from a uniform distribution (Smith et al., 1988) or (the positive part of) a random walk process (Kirchler and Huber, 2007, 2009).

The total amount of cash spending power in the market is[22]

$$C_t = (1 + r_f)^t C_0 + S_0 \sum_{\tau \leq t} d_\tau (1 + r_f)^{t-\tau}, \quad t = 0, \ldots, T. \quad (11)$$

The market liquidity, the maximum number of asset units that can be (potentially) traded (bought or sold), absent any exogenous liquidity (credit) injection or short selling, is[23]

$$L_t = S_0 + C_t, \quad t = 0, \ldots, T. \quad (12)$$

The law of supply and demand in finance reads

$$r_t = \gamma \frac{Z_t}{L_t}, \quad t = 1, \ldots, T. \quad (13)$$

where $\gamma$ is a positive adjustment parameter. This linear price impact holds both in lab (Smith et al., 1988) and field data (Cont et al., 2014), using order flow imbalance as a proxy for excess demand.

The two types of traders are respectively motivated by the following anticipated returns:

$$r_t^i = \begin{cases} \dfrac{p_t^i - p_t}{p_t}, & i \in I, \\ \dfrac{v_t^i - p_t}{p_t}, & i \in II, \end{cases}$$

where $p_t^i$ is a forecast at time $t$ of the asset's future resale price (at some future time $t' > t$) (by a Type I trader), and $v_t^i$ is the asset's fundamental value as forecasted adaptively and recursively (by a type II trader) as follows:

$$v_t^i = (1 + \rho^i)v_{t-1}^i - d_t^i, \quad i \in II, \quad (14)$$

starting from an initial guess $v_0^i$, where $d_t^i$ is a dividend forecast by $i \in II$.

The unitary signed demand (counting supply negatively) for each trading type $i \in I \cup II$ is

$$z_t^i = \mathbf{1}\{r_t^i \geq \rho^i\} - \mathbf{1}\{r_t^i < \rho^i\} = 2\mathbf{1}\{r_t^i \geq \rho^i\} - 1,$$
$$i = 1, \ldots, L_t, \quad t = 1, \ldots, T. \quad (15)$$

where $\mathbf{1}\{\cdot\}$ is an indicator function.[24] We decompose the market liquidity in terms of the two trading Types:

$$L_t = L_t^I + L_t^{II},$$

and write the market excess demand [the sum of the signed elementary demands (15)] in terms of units of market liquidity:

$$\frac{Z_t}{L_t} = \frac{1}{L_t} \sum_{i=1}^{L_t} z_t^i = \frac{L_t^I}{L_t} \frac{\sum_{i \in I} z_t^i}{L_t^I} + \frac{L_t^{II}}{L_t} \frac{\sum_{i \in II} z_t^i}{L_t^{II}}.$$

---

[21] One of the goals of an experimental implementation is to determine the distributions of the relevant variable of the model. Absent any such information, we assume maximum entropy distributions.

[22] Asset sales and revenues being equal in the aggregate, they do not generate any net cash addition.

[23] A more detailed analysis of the endogenous dynamics of market liquidity is not undertaken here. Previous works in this direction include Caginalp et al. (2000).

[24] That is, $\mathbf{1}\{A\} = 1$ if $A$ is true, and $\mathbf{1}\{A\} = 0$, otherwise.

Using an expectation symbol $\mathbf{E}_{i \in J}$ to denote (for simplicity of notation and manipulation) averaging among each trader Type $J$, and using (15) and iterating the expectations, we get:

$$\frac{Z_t}{L_t} = 2\frac{L_t^I}{L_t}\mathbf{E}_{i \in I}[F^I(r_t^i)] + 2\frac{L_t^{II}}{L_t}\mathbf{E}_{i \in II}[F^{II}(r_t^i)] - 1. \tag{16}$$

Eqs. (13) and (16) together with the news-corrected adaptive expectations assumption, defines the general model as follows:

$$d_t = G(d_{t-1}), \quad d_0 = 0, \tag{17}$$

$$r_t = 2\gamma \frac{L_t^I}{L_t^I + L_t^{II}}\mathbf{E}_{i \in I}[F^I(r_t^i)] + 2\gamma \frac{L_t^{II}}{L_t^I + L_t^{II}}\mathbf{E}_{i \in II}[F^{II}(r_t^i)] - \gamma, \tag{18}$$

$$r_t^i = \mu^i r_{t-1}^i + (1-\mu^i)r_{t-1} + \varepsilon_t^i \text{news}_t, \quad 0 \leq \mu^i \leq 1, \quad i \in I, \tag{19}$$

$$d_t^i = \mu^i d_{t-1}^i + (1-\mu^i)d_{t-1} + \varepsilon_t^i \text{news}_t, \quad 0 \leq \mu^i \leq 1, \quad i \in II, \tag{20}$$

$$v_t^i = (1+\rho^i)v_{t-1}^i - d_{t-1}^i, \tag{21}$$

where $\text{news}_t = 1$ if some news arrived in the market in period $t$, and 0, otherwise; $\{\mu^i\}$ are memory parameters, with $0 \leq \mu^i \leq 1$; and $\{\varepsilon_t^i\}$ are the impacts of news on traders' forecasts. The news-corrected adaptive expectations can be written as:

$$r_t^i = (1-\mu^i)\sum_{k=0}^{t-1}(\mu^i)^k r_{t-k} + \sum_{k=0}^{t}(\mu^i)^k \varepsilon_{t-k}^i \text{news}_{t-k}, \quad i \in I, \tag{22}$$

$$d_t^i = (1-\mu^i)\sum_{k=0}^{t-1}(\mu^i)^k d_{t-k} + \sum_{k=0}^{t}(\mu^i)^k \varepsilon_{t-k}^i \text{news}_{t-k}, \quad i \in II, \tag{23}$$

provided $r_0^i = d_0^i = 0$. Perfect memory of past news means $\mu^i = 1$, and zero memory, $\mu^i = 0$.

Absent any knowledge about the minimum acceptable return, we assume it is uniformly distributed in the market:[25]

$$F^J(x) = \text{prob}\{\rho^J \leq x\} = \frac{x}{\rho_{\max}^J}, \quad 0 < x < \rho_{\max}^J, \quad J = I, II. \tag{24}$$

The following variables [Eq. (25)] measure the predominance of each trading type in the market and will play a central role in the theory:[26]

$$n_t^J = 2\frac{\gamma}{\rho_{\max}^J}\frac{L_t^J}{L_t^I + L_t^{II}}, \quad J = I, II. \tag{25}$$

These variables also are not directly observable; we assume they are exponentially distributed.

Finally, assuming the dividend process follows a random walk constrained to be positive, we get the linear version of the general model:

$$d_t = \max\{0, d_{t-1} + \varepsilon_t^d\}, \tag{26}$$

$$p_t = (1+r_t)p_{t-1}, \tag{27}$$

$$r_t = n_t^I r_t^e + n_t^{II}\frac{v_t^e - p_t}{p_t} - \gamma, \tag{28}$$

$$n_t^J = 2\frac{\gamma}{\rho_{\max}^J}\frac{L_t^J}{L_t^I + L_t^{II}}, \quad J = I, II, \tag{29}$$

---

[25] We assume a continuous uniform distribution only for the simplicity of notation and simulation.

[26] A uniform (or other type of) distribution would produce essentially the same result, although the exponential seems to produce returns closer to empirical ones.

$$r_t^e = \mu^I r_{t-1}^e + (1-\mu^I)r_{t-1} + \varepsilon_t^I \text{news}_t, \tag{30}$$

$$d_t^e = \mu^{II} d_{t-1}^e + (1-\mu^{II})d_{t-1} + \varepsilon_t^{II} \text{news}_t, \tag{31}$$

$$v_t^e = (1+\rho)v_{t-1}^e - d_t^e, \tag{32}$$

where $r_t^e = \mathbf{E}_{i \in I}(r_t^i)$, $d_t^e = \mathbf{E}_{i \in II}(d_t^i)$, $v_t^e = \mathbf{E}_{i \in II}(v_t^i)$, $\mu^J = \mathbf{E}_{i \in J}(\mu^i)$, $\varepsilon_t^J = \mathbf{E}_{i \in J}(\varepsilon_t^i)$, $J = I, II$, and $\rho = \mathbf{E}_{i \in II}(\rho^i)$.

## 4. The speculative market model and the power law of returns

The intuition that extreme asset price fluctuations are due to the amplifying feedback inherent to speculative trades can be formally proven in the model (26)–(32) by assuming zero average asset mispricing, so that the asset return is driven by Type I traders:

$$v^e = p \text{ and } 0 \leq \mu^I < 1, \tag{33}$$

which yields the speculative return model:

$$r_t = n_t^I r_t^e - \gamma, \tag{34}$$

$$n_t^I = 2\frac{\gamma}{\rho_{\max}^I}\frac{L_t^I}{L_t^I + L_t^{II}}, \tag{35}$$

$$r_t^e = (1-\mu)\sum_{k=0}^{t-1}\mu^k r_{t-k} + \sum_{k=0}^{t}\mu^k \varepsilon_{t-k}\text{news}_{t-k}. \tag{36}$$

In the simplest case of purely speculative market where traders have zero memory of exogenous news, we get the first order random-coefficient autoregressive return process:

$$v^e = p \text{ and } \mu^I = 0 \text{ implies } r_t = n_t^I r_{t-1} + n_t^I \varepsilon_t^I \text{news}_t - \gamma. \tag{37}$$

The general case $v^e = p$ and $0 < \mu < 1$ is qualitatively equivalent (as regards tail behaviors) to the simple case (37). Combining (34) and (36) yields indeed the expanded form of the speculative return process:

$$r_t = n_t^I \sum_{k=0}^{t-1}(1-\mu^I)(\mu^I)^k r_{t-k} + n_t^I \sum_{k=0}^{t}(\mu^I)^k \varepsilon_{t-k}\text{news}_{t-k} - \gamma. \tag{38}$$

Assume the (random) limit $r_\infty = \lim_{t \to \infty} r_t$ exists (almost surely), and likewise for the other random variables involved in (38), which is the case if these latter are i.i.d. (independent and identically distributed), as we will assume throughout. Then $r_\infty$ is the stationary solution to the stochastic recurrence equation (38) and obeys the equality in distribution[27]

$$r_\infty \stackrel{d}{=} n_\infty^I \sum_{k=0}^{\infty}(1-\mu^I)(\mu^I)^k r_\infty + n_\infty^I \sum_{k=0}^{\infty}(\mu^I)^k \varepsilon_\infty \text{news}_\infty - \gamma,$$

which simplifies to

$$r_\infty \stackrel{d}{=} n_\infty^I r_\infty + (1-\mu^I)^{-1} n_\infty^I \varepsilon_\infty \text{news}_\infty - \gamma. \tag{39}$$

This is then the stationary solution (unique in distribution) of the stochastic recurrence equation (38), and it has power-law tails, according to the following result, where $e_t^I = \varepsilon_t^I \text{news}_t$.

**Proposition** (*Power Law of Speculative Returns*). Consider the linear model with $p = v^e$ and $0 < \mu^I < 1$. Assume $\{(n_t^I, e_t^I)\}$ are i.i.d. copies of a random pair $(n^I, e^I)$ whose moments are all finite, $e^I(1-n^I)^{-1}$ is non-degenerate (nonconstant), and the law of $\log(n^I)$ given $n^I > 0$ is nonarithmetic.[28] If there is $\alpha > 0$ such that $\mathbf{E}(n^I)^\alpha = 1$ then prob$\{|r_\infty| \geq x\} \sim cx^{-\alpha}$, $c > 0$.

---

[27] An equality in distribution means the left-hand side and right-hand side have the same distribution.

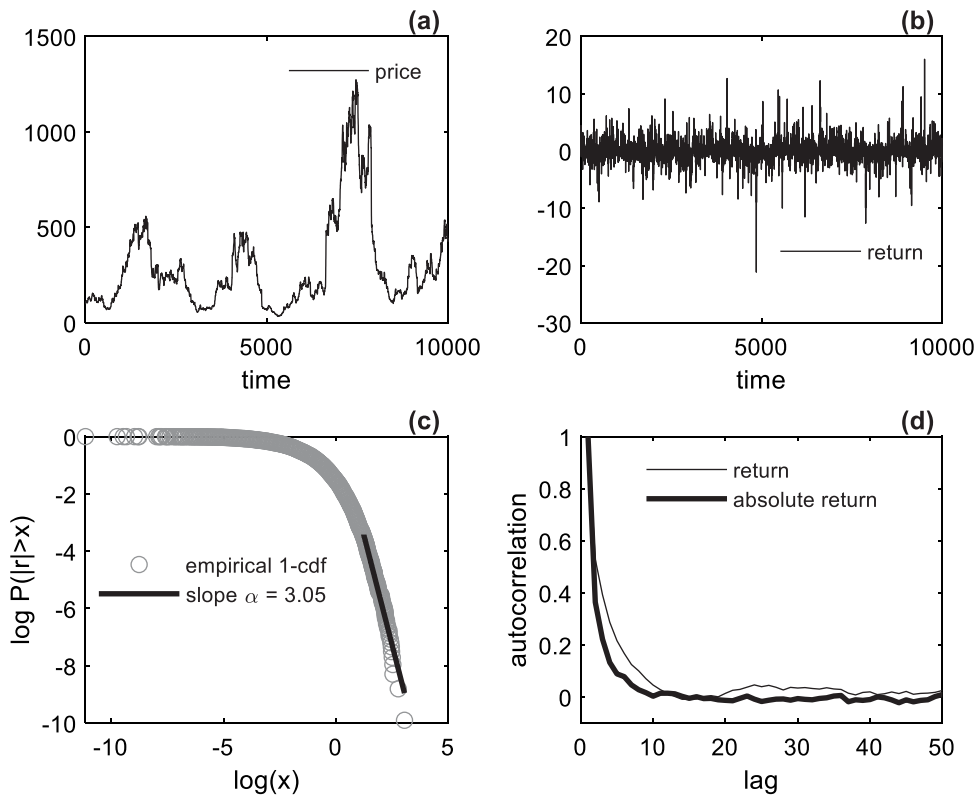

**Fig. 6.** The speculative asset market model simulated: prob{news} = 1, disposition to speculate $\{n_t^I\}$ i.i.d. draws from an exponential distribution with mean 0.55; impact of news $\{\varepsilon_t^I\}$ is zero-mean Gaussian with standard deviation of 1.

**Proof.** This is an implication of Kesten theorem (Buraczewski et al., 2016, Theorem 2.4.4) stated formally and explained intuitively in Appendix. Moreover, one can show by adapting standard stationarity analysis of RCAR processes (Buraczewski et al., 2016, Section 2.1) that $0 < \mu < 1$, $\mathbf{E}(n^I)^\alpha = 1$, and the moment condition of the proposition imply that (39) is indeed the stationary solution of the stochastic recurrence equation (38), and the return process converges in distribution to the stationary solution.[29] ∎

The speculative market model is simulated in Fig. 6.

As mentioned in Section 2 [Eq. (9)] and as is clear from Fig. 6, the speculative market model cannot account for clustered volatility: for any such autoregressive model, and for any arbitrary function $f$, cov$[f(r_{t-h}), f(r_t)]$, when it is well-defined, decays rapidly (at an exponential rate) with the lag $h$ (Mikosch and Starica, 2000; Basrak et al., 2002). So, volatility, whether measured as $|r|$, $r^2$, or more generally by any function $f$, cannot be long-range correlated in this speculative model. The key to this model's incapacity for reproducing clustered volatility is the short-memory property: speculators forget fundamental news very quickly, at the exponential rate $(\mu^I)^h$, as is clear from (38), and this short memory forbids any persistent trading behavior capable of explaining a persistent volatility. Setting $\mu^I \approx 1$ in the speculative model would lead to long memory of both the return and the absolute return, however, unlike in empirical data. A more general model is therefore needed involving the long memory of news in both trading types, whose interplay yields the stylized facts.

### 5. A more general specification

It turns out that both the fat tails and the clusters of volatility are robustly captured by the linear model (26)–(32) assuming the specification:

$$\mu^I = \mu^{II} = 0.99, \qquad (40)$$

as the simulation in Fig. 7 shows, where the other parameter specifications, reported in Table 1, are chosen to have realistic orders of magnitude compared to empirical data (notably the standard deviation of return which is typically around 1% per day in field data.) (See Fig. 8.)

Excess volatility of price relatively to fundamental value is also a generic phenomenon in the model. To investigate the bubble phenomenon in the context of the bubble-and-crash experiments, we specialize the model's asset present value process $\{v_t\}$ to the standard experimental implementation, by drawing dividends randomly and uniformly from the set $\{0, 4, 8, 20\}$, and setting $\rho = 0$. According to the model simulations, the asset bubble size increases ceteris paribus with the dominance of speculation in the market, namely $\mathbf{E}(n^I)$, and it decreases with the overall trading horizon $T$, as shows Fig. 10. Also, the traders' adaptive expectation of the fundamental value being close to the neoclassical version $v_t^e = (T-t+1)\mathbf{E}(d_t) \to 0$, we recover the power law tail exponent $\alpha = 1$ as explained in Section 2 and confirmed in Fig. 10. (See Fig. 9.)

---

[28] A random variable is nonarithmetic if its support does not coincide with integer multiples of a real number.

[29] The proposition assumes a less general setting to avoid more general but more technical assumptions on $(n, \varepsilon)$, met by any variables with finite moments. By Jensen's inequality, $\mathbf{E}(n^\alpha) = 1$ implies $\mathbf{E}[\log n] < 0$, which, together with the moment assumption, guarantees the existence of the stationary solution: the latter condition generalizes the known one, $n < 1$, for a nonrandom $n$. (In these statements, absolute values would be considered instead, were the variables nonpositive.) For an exponential distribution $n^I$, one can show that $\mathbf{E}(n^I)^3 = 1$ for $\mathbf{E}(n^I) \approx 0.55$ as assumed in Fig. 6.

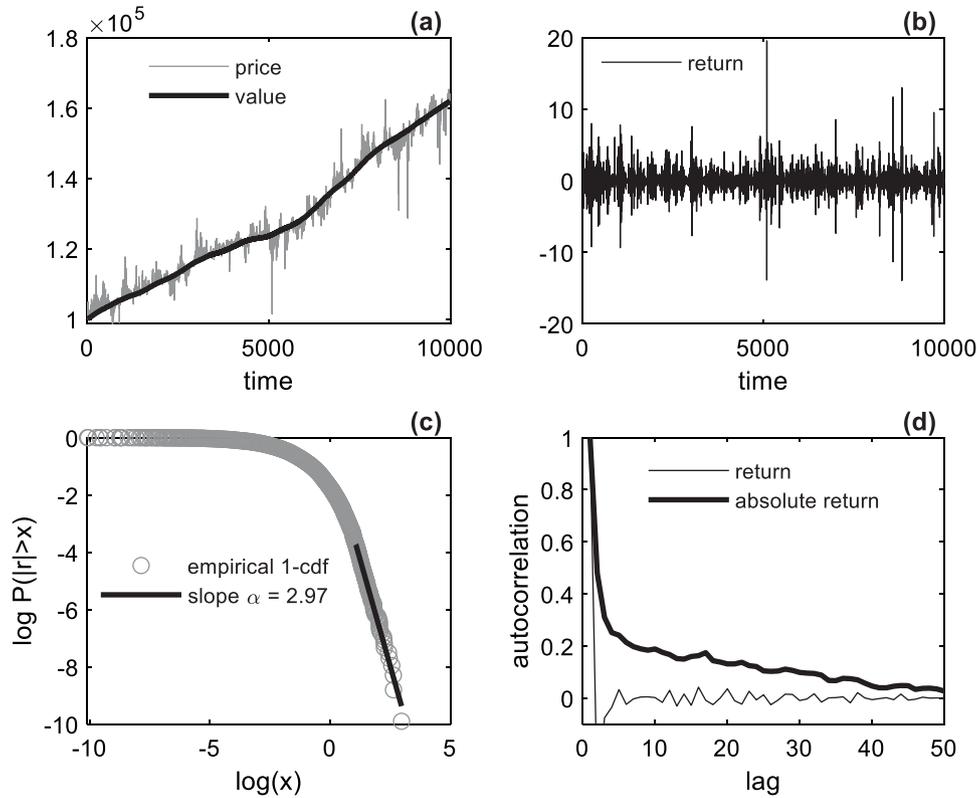

**Fig. 7.** The general model simulated: (a) price; (b) return (in percent); (c) cumulative distribution of volatility in log–log scale, and a linear fit of the tail; (d) autocorrelation function of return and absolute return.

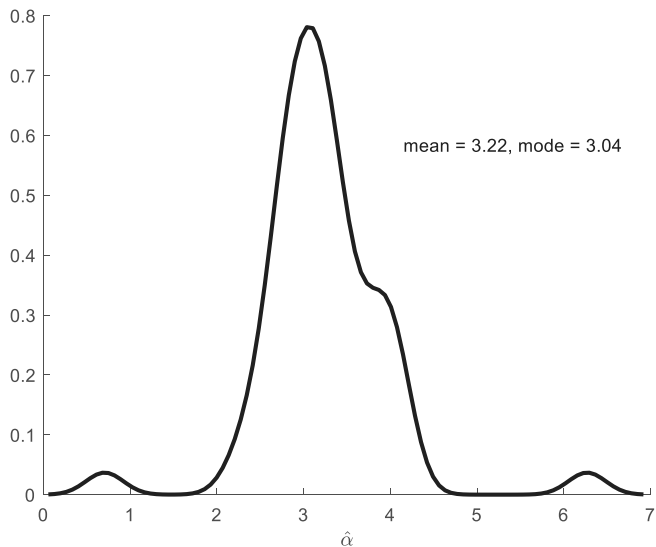

**Fig. 8.** Density function estimate for the tail exponent estimator $\hat{\alpha}$ for 50 sample paths of the model with the parameters of Table 1 (General Model).

**Table 1**
Model parameter specification and summary statistics. Simulations with $T = 10000$; initial conditions: $p_0 = v_0^e = 10^5$, $d_0 = d_0^e = 10$, and $r_1 = r_1^e = 0$. The $\{n_t^I\}$ are random draws from exponential distributions; all $\varepsilon$ variables are zero-mean normally distributed, except in the speculative model, where mean $(\varepsilon^I) = 0.01\%$.

|  | Fig. 6 Speculative model | Fig. 7 General model |
|---|---|---|
| **Parameters** | | |
| Discount rate $\rho$ | N/A | 1.64% |
| Price impact: $\gamma$ | 0.01% | 0.01% |
| Dividend: std($\varepsilon^d$) | N/A | 0.1 |
| mean($n^I$) | 0.55 | 0.2 |
| mean($n^{II}$) | N/A | 0.8 |
| News: std($\varepsilon^I$) | 1% | 1% |
| News: std($\varepsilon^{II}$) | N/A | 0.1 |
| prob(news) | 1 | 0.2 |
| Memory $\mu^I$ | 0.99 | 0.99 |
| Memory $\mu^{II}$ | N/A | 1 |
| mean($r$) | 0.01% | 0.01% |
| std($r$) | 1.31% | 1.06% |
| Tail exponent $\hat{\alpha}$ | 3.05 | 2.97 |

## 6. Summary and leads for future work

In this paper we set the foundations for a classical model of financial markets to explain three of the most general stylized facts of speculative prices: the fat tails of speculative returns emerge inherently from speculation under adaptive expectations; the clustered volatility is a consequence of the long memory that traders have of news; and bubbles are an inherent phenomenon whose size increases with the preponderance of speculation in the market and decreases with the trading horizon, ceteris paribus.

Some aspects of the model call for further investigation. For example, although the simulations suggest fat-tailed returns in the general specification of the model, the exact tail behavior needs to be investigated mathematically; also, an experimental investigation of the model would be desirable, to know, among other things, the distribution of the key variables (the minimum acceptable return, proportion of speculation, the impact of news). The endogenous dynamics of market liquidity remains to be analyzed. And, moreover, the theoretical discussion of the bubble-and-crash phenomenon, as replicated by the model, is of course too stylized in view of the rich and subtle laboratory studies of this phenomenon; in particular, we have not yet modeled a crucial aspect of lab market dynamics in general, cross-period subject learning and experience.

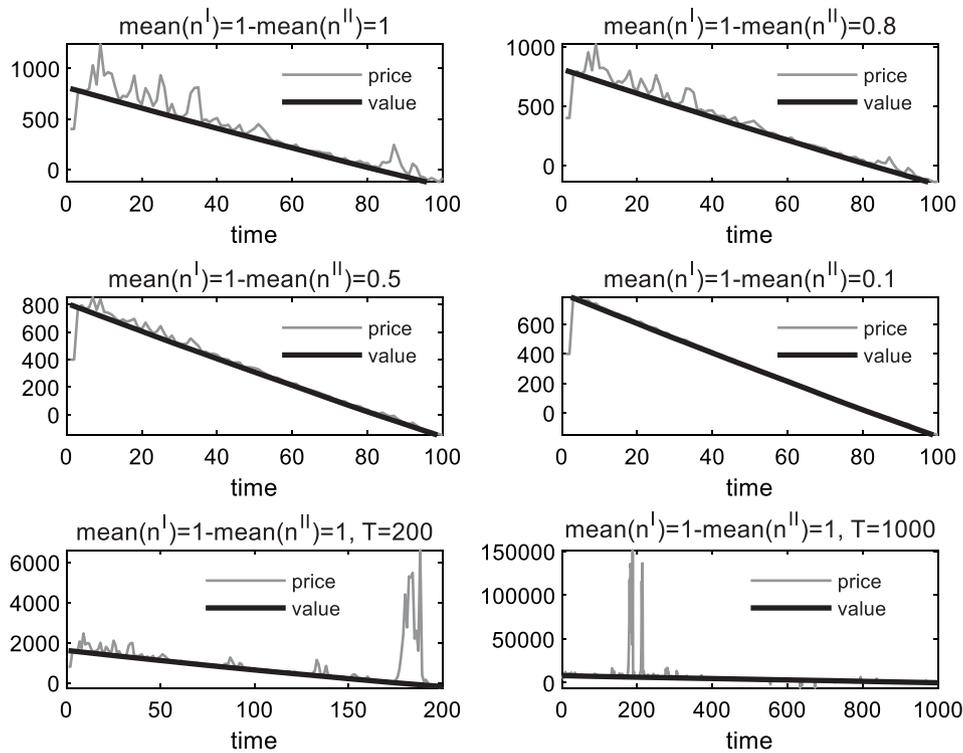

**Fig. 9.** Bubble and crash phenomena in the model: the size of the bubble is increasing with the propensity to speculate in the market and the time horizon, ceteris paribus. The model parameters are as in Table 1, General Model, save: dividends uniformly drawn from $\{0, 4, 8, 20\}$; $\rho = 0$; and $p_0 = v_0^e/2 = 400$.

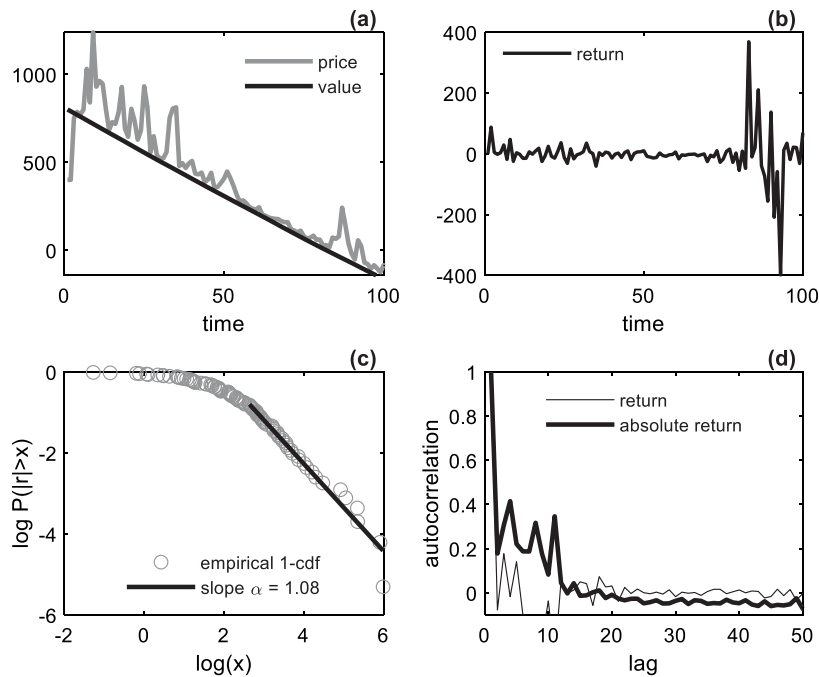

**Fig. 10.** With a present value declining to zero, the model recovers the power exponent $\alpha \approx 1$, as expected. All the model parameters are as in Table 1, General Model, save: the dividends, uniformly drawn from the set $\{0, 4, 8, 20\}$; $T = 100$; $p_0 = v_0^e/2 = 400$; and $\rho = 0$.

**Appendix. Kesten theorem**

We offer an intuitive explanation for the following theorem due to Kesten (1973) and proven differently by Goldie (1991) and thus also known as the Kesten–Goldie theorem; see also Vervaat (1979), among others; for a detailed exposition, see Buraczewski et al. (2016, Section 2.4.4).

**Theorem.** *Let $(a, b)$ be a pair of real-valued random variables. Assume there is $\alpha > 0$ such that:* $\mathbf{E}(|a|^\alpha) = 1$, $\mathbf{E}[|a|^\alpha \max(0, \log|a|)] < \infty$, *and* $\mathbf{E}(|b|^\alpha) < \infty$. *Assume $\log|a|$, given $a \neq 0$, is nonarith-* metic, and $(1-a)^{-1}b$ is nondegenerate (nonconstant). Then $r \stackrel{d}{=} ar + b$ implies prob$\{r > x\} \sim c_+ x^{-\alpha}$, prob$\{r < -x\} \sim c_- x^{-\alpha}$, $x \to \infty$, hence prob$\{|r| > x\} = cx^{-\alpha}$, $c_\pm \geq 0$, $c = c_+ + c_- > 0$.

The equation in law

$$r \stackrel{d}{=} ar + b \quad (A)$$

arises, as we saw in the text, as the stationary solution of a RCAR process, say $r_t = a_t r_{t-1} + b_t$, where $(a_t, b_t)$ are independent copies of a generic random pair $(a, b)$. One can think of such process in various contexts as modeling the state of a system driven by an exogenous influence $b$ amplified by an endogenous feedback term $ar$. A fat-tailed output $r$ can of course result directly from a fat-tailed input $b$ (Grincevićius, 1975; Grey, 1994). But, more surprisingly, as Kesten originally proved, a fat-tailed, power-law, output $r$ can emerge from a light–tail pair $(a, b)$ through feedback amplifications. (For our purpose, a variable is fat-tailed if one of its moments is infinite; power laws are important examples.)

We are interested in the tail behavior of the distribution of the output $r$, finding a function

$$H(x) \sim \text{prob}\{|r| \geq x\}, \quad x \to \infty.$$

Provided $ar \neq 0$, we can write (A) as follows:

$$|r| \stackrel{d}{=} |ar|(1 + |b/ar|). \quad (B)$$

Assuming $b$ is "relatively light-tailed", $|b/ar|$ is "negligible" for big realizations of $r$, hence we have prob$\{|r| \geq x\} \sim \text{prob}|r| > x/|a|$, and by iterated expectations (integrating across the distribution of $a$) one gets the functional equation

$$H(x) = \mathbf{E}_a[H(x/|a|)]. \quad (C)$$

For real numbers $x$ and $y \neq 0$, assume there is a function $h(y)$ such that[30]

$$H(x/y) = H(x)h(y). \quad (D)$$

Set $y = |a|$ in (D) to get $H(x/|a|) = H(x)h(|a|)$, which in view of (C) implies

$$\mathbf{E}_a[h(a)] = 1. \quad (E)$$

Set $x = 1$ in (D) to get $H(1/y) = H(1)h(y)$, hence

$$h(y) = H(1/y)/H(1). \quad (F)$$

Thus (D) becomes $H(x/y) = H(x)H(1/y)/H(1)$, which in terms of $z = 1/y$ and $h(x) = H(x)/H(1)$ reads

$$h(xz) = h(x)h(z). \quad (G)$$

The only continuous solution to (G) is the power law $h(x) = Cx^{-\alpha}$, so

$$H(x) = H(1)x^{-\alpha}. \quad (H)$$

Going backward, we get, from (F), $h(|a|) = |a|^\alpha$, and (E) becomes

$$\mathbf{E}(|a|^\alpha) = 1. \quad (I)$$

To summarize the steps:

1. The power law $H(x) = H(1)x^{-\alpha}$ holds asymptotically: prob$\{|r| \geq x\}/H(x) \to 1$, $x \to \infty$.
2. The feedback component is the crucial one in the emergence of the power law, since $b/ar$ is assumed small.
3. The crucial step assumes existence of the function $h$, in particular existence of $\alpha > 0$ such that $\mathbf{E}_a[h(a)] = \mathbf{E}(|a|^\alpha) = 1$, a necessary condition of which is prob$\{|a| > 1\} > 0$ (otherwise $\mathbf{E}(|a|^\alpha) < 1$). Thus the emergence of the power law is due to the amplifying feedbacks, namely events $\{a > 1\}$. Interestingly the power law tail is determined entirely by the feedback variable $a$, through the equation $\mathbf{E}(|a|^\alpha) = 1$.

---

[30] Needless to say, the crucial step is precisely the justification of (D): the ongoing exercise is merely intended to demystify the emergence of the important Eq. (I), rather than to sketch a proof.